\documentclass{article}

\usepackage[english]{babel}
\usepackage[letterpaper,top=2cm,bottom=2cm,left=3cm,right=3cm,marginparwidth=1.75cm]{geometry}
\usepackage{siunitx}
\usepackage{amsmath}
\usepackage{graphicx}
\usepackage[colorlinks=true, allcolors=blue]{hyperref}
\usepackage{subcaption}
\usepackage{xcolor}

\title{Sensitivity Enhancement of S-Band Rydberg Atom Microwave Receiver Using Resonant Cavity}

\author{Yipeng Xie$^{1,2}$, Xinbing Chen$^{3}$, Mingwei Lei$^{1}$, Meng Shi$^{1,*}$}

\begin{document}
\date{}  
\maketitle

{\small

\noindent
$^{1}$ Key Laboratory of Space Utilization, Technology and Engineering Center for Space Utilization, Chinese Academy of Sciences, 100094, Beijing, China\\
$^{2}$ University of Chinese Academy of Sciences, 100049, Beijing, China\\
$^{3}$ Beijing University of Posts and Telecommunications, 102206, Beijing, China
\noindent\\[2mm]
\textbf{*Corresponding author:} \texttt{shimeng@csu.ac.cn}
}

\vspace{5mm}

\begin{abstract}
Rydberg atom-based microwave electric field sensing has attracted growing interest owing to its inherent advantages, such as absolute calibration, wideband operability, and compatibility with room-temperature devices. A critical bottleneck that limits sensitivity is the inefficient coupling between the Rydberg atoms and the incident microwave field, particularly when detecting weak signals propagating in free space. Here we propose and experimentally validate a scheme that integrates a horn antenna with a resonant microwave cavity to significantly improve this coupling for free-space signal reception in the S-band. Using a two-photon excitation scheme in a cesium vapor cell, we systematically characterize the sensing performance under three configurations: a bare cell, direct cavity injection, and a cavity coupled to a horn antenna that captures free-space microwave signals over a 1 m distance. In the antenna-coupled cavity configuration, we achieve an optimal sensitivity of 2.33 nV/cm/$\sqrt{\text{Hz}}$ at the receiving antenna, which corresponds to an enhancement of approximately 17.9 dB compared to the optimized bare vapor cell configuration. Our findings offer a practical and effective route to boost the sensitivity of Rydberg atomic sensors, facilitating their adoption in real-world microwave metrology and wireless communication applications where weak free-space electric fields must be reliably measured.
\end{abstract}

\section{Introduction}
Detecting microwave fields is important for applications like communications\cite{Rappaport2013,Saad2020}, radar\cite{Fan2025,Tan2025}, and precision sensing\cite{Huang2025,Sun2025}. Rydberg atom-based sensors have shown great potential in these areas in recent years\cite{Robinson2021,Holloway2019,Holloway2019-2,Jiao2019,Meyer2018}. Consequently, they have emerged as a promising platform for electric field metrology, offering absolute, SI-traceable measurements with high sensitivity across a broad frequency range in compact, room-temperature systems\cite{Schlossberger2024}. Since the sensing properties of Rydberg atoms were first explored in 1980 \cite{Figger1980}, various methods have been employed to improve receiver sensitivity, including laser repumping \cite{Prajapati2021}, parameter optimization \cite{Cai2022}, and N-wave mixing \cite{Borowka2024}.However, the central challenge lies in how to achieve effective coupling between the Rydberg atoms and the microwave electric field\cite{Cox2021,Yang2023,Holloway2022,Walker2022,Hogan2012,Morgan2020}.Conventional MW antennas achieve directional gain by collecting the MW electric field within a specific directional range. In contrast, the sensing element of a Rydberg atomic measurement system is a glass vapor cell, which lacks directional gain. Moreover, the external MW field tends to disperse in free space, and the region within the vapor cell where Rydberg atoms exist is very limited due to excitation laser constraints. This makes effective coupling between the Rydberg atoms and the external MW field difficult, resulting in significant insertion loss and substantially reduced sensitivity.

This paper addresses the challenge of receiving weak free-space microwave signals in the S-band by proposing a novel detection scheme for Rydberg atoms using a horn antenna coupled with a resonant cavity. Actually, recent studies have explored Rydberg-atom-based satellite signal reception using microwave resonator cavities with front-end antennas, including soil moisture remote sensing via signals of opportunity~\cite{Arumugam2024} and, as demonstrated in our team's previous work, direct C-band satellite beacon detection with a cavity-coupled receiver that operates without low-noise amplifier(LNA)~\cite{Lei2026}, a key distinction from the Jet Propulsion Laboratory(JPL) approach which relies on LNA for signal extraction. In contrast to existing resonant cavity-enhanced works~\cite{Sandidge2024,Liu2025}, which are limited to direct injection or near-field coupling, this scheme takes advantage of both the directional gain of the horn antenna and the cavity's mode-matching property, which generates a standing wave and sustained oscillation at the target frequency, leading to a strong enhancement of the local microwave electric field. This study achieves stable and high-sensitivity detection over a long-distance wireless link of 1 m. Compared to the optimized bare vapor cell baseline which achieves its best sensitivity through careful optimization of laser beam sizes and cell temperature, the proposed scheme combining antenna gain and cavity enhancement yields a sensitivity improvement of 17.9 dB, reaching an optimal sensitivity of 2.33 nV/cm/$\sqrt{\text{Hz}}$.
This work overcomes the confined laboratory scenario limitations of conventional cavity-enhanced methods, and the achieved sensitivity surpasses that of most other reported sensitivity enhancement approaches for Rydberg-atom-based microwave sensors~\cite{Tu2024,Borowka2024,Sandidge2024,Yan2025,Wu2025,Ding2022}, providing a viable pathway for the application of such sensors in real-world wireless communications. It is expected to promote their practical development in areas such as radar detection, wireless communication and electromagnetic spectrum sensing.

\section{Theory}
\begin{figure}[t]
    \centering
    \includegraphics[width=1\linewidth]{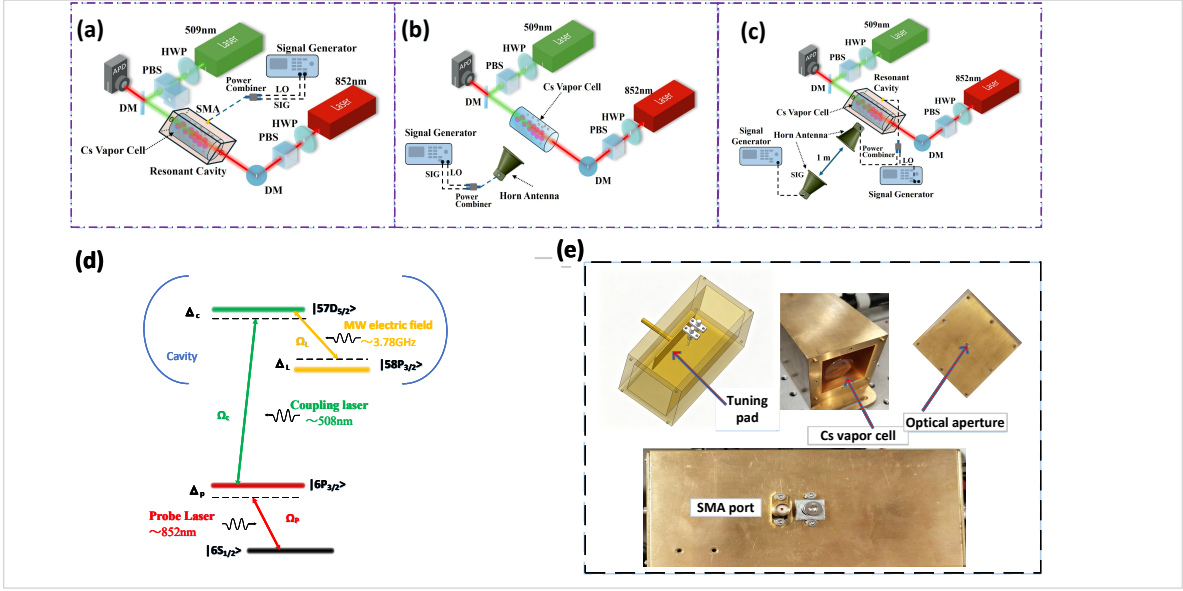}
    \caption{Schematic of experimental setup. (a) Direct cavity injection configuration. (b) Free-space (bare cell) configuration. (c) Antenna-coupled cavity configuration. (d) Energy-level diagram of two-photon Rydberg-EIT. (e) Photograph of the fabricated microwave resonant cavity.}
    \label{fig:exp_setup}
\end{figure}

The Rydberg atom-based microwave receiver operates on the principle of electromagnetically induced transparency (EIT) and Autler-Townes (AT) splitting. As illustrated in Fig.~\ref{fig:exp_setup}(d), we consider a four-level atomic system consisting of a ground state $|1\rangle $, an intermediate excited state $|2\rangle $, a Rydberg state $|3\rangle $, and an adjacent Rydberg state $|4\rangle$. The probe laser at 852\,nm drives the transition $|1\rangle \rightarrow |2\rangle$ with Rabi frequency $\Omega_{\mathrm{p}}$, while the coupling laser at 509\,nm drives the transition $|2\rangle \rightarrow |3\rangle$ with Rabi frequency $\Omega_{\mathrm{c}}$. The microwave field to be detected drives the Rydberg transition $|3\rangle \rightarrow |4\rangle$ with Rabi frequency $\Omega_{\mathrm{MW}}$.Under the rotating wave approximation, the Hamiltonian describing this four-level system interacting with the optical and microwave fields can be written as:
\begin{equation}
H = \hbar
\begin{pmatrix}
0 & \frac{\Omega_{\mathrm{p}}}{2} & 0 & 0 \\[6pt]
\frac{\Omega_{\mathrm{p}}}{2} & -\Delta_{\mathrm{p}} & \frac{\Omega_{\mathrm{c}}}{2} & 0 \\[6pt]
0 & \frac{\Omega_{\mathrm{c}}}{2} & -(\Delta_{\mathrm{p}} + \Delta_{\mathrm{c}}) & \frac{\Omega_{\mathrm{MW}}}{2} \\[6pt]
0 & 0 & \frac{\Omega_{\mathrm{MW}}}{2} & -(\Delta_{\mathrm{p}} + \Delta_{\mathrm{c}} - \Delta_{\mathrm{MW}})
\end{pmatrix}
\end{equation}
where $\Delta_{\mathrm{p}} = \omega_{\mathrm{p}} - \omega_{12}$ is the probe laser detuning, $\Delta_{\mathrm{c}} = \omega_{\mathrm{c}} - \omega_{23}$ is the coupling laser detuning, $\Delta_{\mathrm{MW}} = \omega_{\mathrm{MW}} - \omega_{34}$ is the microwave detuning, and $\omega_{ij}$ represents the resonant frequency of the $|i\rangle \rightarrow |j\rangle$ transition. The Rabi frequencies are defined as $\Omega_{\mathrm{p}} = \boldsymbol{\mu}_{12} \cdot \mathbf{E}_{\mathrm{p}}/\hbar$, $\Omega_{\mathrm{c}} = \boldsymbol{\mu}_{23} \cdot \mathbf{E}_{\mathrm{c}}/\hbar$, and $\Omega_{\mathrm{MW}} = \boldsymbol{\mu}_{34} \cdot \mathbf{E}_{\mathrm{MW}}/\hbar$, where $\boldsymbol{\mu}_{ij}$ is the transition dipole moment and $\mathbf{E}$ represents the respective electric field amplitudes.The dynamics of the atomic system are governed by the Lindblad master equation for the density matrix $\rho$:
\begin{equation}
\dot{\rho} = \frac{\partial \rho}{\partial t} = -\frac{i}{\hbar}[H, \rho] + \mathcal{L}[\rho]
\end{equation}
where $\mathcal{L}[\rho]$ is the Lindblad superoperator accounting for atomic decay and decoherence processes:
\begin{equation}
\mathcal{L}[\rho] = \sum_{k} \left( L_k \rho L_k^{\dagger} - \frac{1}{2} \{L_k^{\dagger} L_k, \rho\} \right)
\end{equation}
The collapse operators $L_k$ include the spontaneous decay from the intermediate state $|2\rangle$ with rate $\Gamma_2$, the decay from the Rydberg states $|3\rangle$ and $|4\rangle$ with rates $\Gamma_3$ and $\Gamma_4$, as well as the dephasing rates $\gamma_{ij}$ between levels $|i\rangle$ and $|j\rangle$ arising from laser linewidth, atomic collisions, and transit time broadening.

In the superheterodyne detection configuration, the microwave field incident on the Rydberg atoms consists of two components: a strong local oscillator (LO) field and a weak signal (SIG) field. The total microwave electric field can be expressed as:
\begin{equation}
E_{\mathrm{MW}}(t) = E_{\mathrm{LO}} \cos(\omega_{\mathrm{LO}} t) + E_{\mathrm{SIG}} \cos(\omega_{\mathrm{SIG}} t + \phi_{\mathrm{SIG}})
\end{equation}
where $E_{\mathrm{LO}}$ and $E_{\mathrm{SIG}}$ are the field amplitudes, $\omega_{\mathrm{LO}}$ and $\omega_{\mathrm{SIG}}$ are the angular frequencies, and $\phi_{\mathrm{SIG}}$ is the relative phase of the signal field. Under the condition $E_{\mathrm{SIG}} \ll E_{\mathrm{LO}}$, the total Rabi frequency can be approximated as:
\begin{equation}
\Omega_{\mathrm{MW}}(t) = \Omega_{\mathrm{LO}} \cos(\omega_{\mathrm{LO}} t) + \Omega_{\mathrm{SIG}} \cos(\omega_{\mathrm{SIG}} t + \phi_{\mathrm{SIG}})
\end{equation}
with $\Omega_{\mathrm{LO}} = \mu_{34} E_{\mathrm{LO}}/\hbar$ and $\Omega_{\mathrm{SIG}} = \mu_{34} E_{\mathrm{SIG}}/\hbar$. When the LO field is resonant with the Rydberg transition ($\omega_{\mathrm{LO}} = \omega_{34}$) and the SIG field is slightly detuned by $\delta = \omega_{\mathrm{SIG}} - \omega_{\mathrm{LO}}$, the interference between these two fields produces a beat note in the atomic response. The probe transmission through the atomic vapor cell exhibits an oscillatory component at the difference frequency $\delta$:
\begin{equation}
P_{\mathrm{out}}(t) = P_0 + \kappa \Omega_{\mathrm{SIG}} \cos(2\pi \delta t + \phi_{\mathrm{SIG}})
\end{equation}
where $P_0$ is the steady-state probe transmission, and $\kappa$ is a linear expansion coefficient that depends on the atomic parameters and the LO field strength. This heterodyne detection scheme effectively down-converts the microwave signal to an intermediate frequency $\delta$ in the kHz to MHz range, where it can be conveniently measured using conventional electronic spectrum analyzers. 

In the free-space configuration, the microwave field incident on the atomic vapor cell is inherently weak due to wave divergence and the limited interaction volume. Integrating the vapor cell with a microwave resonant cavity addresses this limitation through spatial confinement of the electromagnetic mode and resonant field amplification. In the present experiment, the cell is positioned at the cavity center, where the electric field reaches its maximum, ensuring that the Rydberg atoms interact with the strongest available field.The field buildup inside a resonant cavity is quantified by the quality factor, defined as \(Q = \omega_0 U / P_{\mathrm{loss}}\), where \(U\) is the stored electromagnetic energy and \(P_{\mathrm{loss}}\) the dissipated power. On resonance, the stored energy scales with the incident power as \(U \propto Q P_{\mathrm{inc}}\). Since the cavity electric field amplitude satisfies \(|E_{\mathrm{cav}}|^{2} \propto U\), it follows that \(|E_{\mathrm{cav}}| \propto \sqrt{Q} \, |E_{\mathrm{inc}}|\). Incorporating impedance matching and insertion losses into a coupling coefficient \(\beta\), the steady-state field at the cavity center, \(E_{\mathrm{cav}}\), is related to the incident field at the feed port, \(E_{\mathrm{inc}}\), by
\begin{equation}
E_{\mathrm{cav}} = \beta \sqrt{Q_{\mathrm{eff}}} \, E_{\mathrm{inc}},
\label{eq:field_enhance}
\end{equation}
where \(Q_{\mathrm{eff}}\) is the effective quality factor, equal to the loaded quality factor \(Q_{L}\) at zero frequency detuning. Equation~(\ref{eq:field_enhance}) shows that the cavity acts as a coherent field amplifier: for a given incident power, the local field experienced by the atoms is enhanced by a factor of \(\beta\sqrt{Q_{\mathrm{eff}}}\) relative to the free-space value.

The sensitivity of a Rydberg-atom microwave receiver is set by the minimum microwave field that produces a resolvable Autler–Townes splitting, corresponding to a minimum detectable Rabi frequency \(\Omega_{\mathrm{MW}}^{\mathrm{min}}\). Because the atoms are placed at the field antinode in both configurations, the threshold local field at the atomic position is identical. In free space, the local field equals the incident field, giving \(E_{\mathrm{free}}^{\mathrm{min}} = \hbar\Omega_{\mathrm{MW}}^{\mathrm{min}}/\mu_{34}\). In the cavity configuration, the local field is amplified according to Eq.~(\ref{eq:field_enhance}), so that a much weaker incident field \(E_{\mathrm{cavity}}^{\mathrm{min}}\) suffices. Equating the required local fields yields the sensitivity enhancement factor:
\begin{equation}
\frac{E_{\mathrm{free}}^{\mathrm{min}}}{E_{\mathrm{cavity}}^{\mathrm{min}}} = \beta \sqrt{Q_{\mathrm{eff}}}.
\label{eq:sensitivity_enhance}
\end{equation}
In logarithmic units, the sensitivity improvement is \(20\log_{10}(\beta\sqrt{Q_{\mathrm{eff}}})\;\mathrm{dB}\) theoretically.

\section{Experiment Setup}
\begin{figure}[t]
    \centering
    \includegraphics[width=1\linewidth]{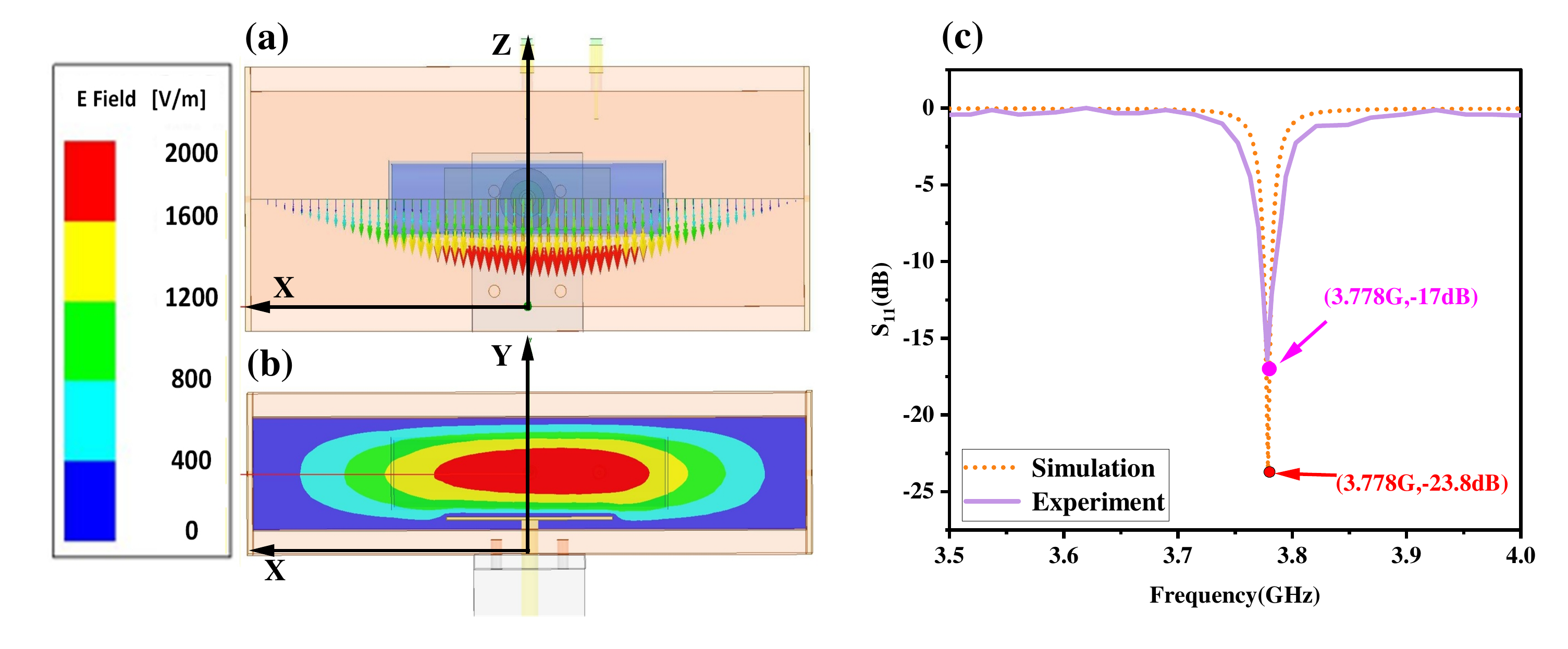}
    \caption{Electromagnetic simulations of the microwave resonant cavity performed with HFSS. (a) Electric field distribution inside the cavity, showing strong concentration at the central region where the atomic vapor cell is placed. (b) Directional uniformity of the electric field across the cavity, confirming that the polarizations of the probe laser, coupling laser, and the microwave field remain aligned for efficient optical readout. (c) Frequency response of the $|S_{11}|$ parameter around 3.8 GHz, where a deep resonance dip of $-23.8$\,dB at 3.778\,GHz is observed in the simulation. In the actual experiment, the measured $S_{11}$ parameter was $-17$\,dB.}
    \label{HFSScavity}
\end{figure}
 Fig.~\ref{fig:exp_setup}(d) presents the atomic energy level structure. A two-photon excitation method is employed to transfer atoms from the ground state to the Rydberg state. A probe laser at 852\,nm drives the \(|6S_{1/2}\rangle \rightarrow |6P_{3/2}\rangle\) transition while a coupling laser at 509\,nm drives the \(|6P_{3/2}\rangle \rightarrow |57D_{5/2}\rangle\) transition. A microwave field drives the \(|57D_{5/2}\rangle \rightarrow |58P_{3/2}\rangle\) . In the experimental arrangement shown in Fig.~\ref{fig:exp_setup}, the probe laser traverses the atomic vapor cell, and the coupling laser propagates counter to the probe laser direction. The vapor cell is a cylindrical cell with a length of 10\,cm and a diameter of 3\,cm. For the two cavity-based configurations (Fig.~\ref{fig:exp_setup}(a) and~\ref{fig:exp_setup}(c)), the probe laser is focused inside the cell to a \(1/e^2\) waist radius of approximately 900\,\(\mu\)m, and the coupling laser is focused to a \(1/e^2\) waist radius of about 1000\,\(\mu\)m. Both measurements are carried out at room temperature (25\,\(^\circ\)C). For the bare vapor cell configuration (Fig.~\ref{fig:exp_setup}(b)), parameter optimization is performed to optimize the electric field sensitivity. In this case, the probe laser waist radius is increased to 1400\,\(\mu\)m, the coupling laser waist radius is increased to 1500\,\(\mu\)m. The probe laser is finally detected by an avalanche photodiode (APD) for signal measurement.

The distinction between Fig.~\ref{fig:exp_setup}(b) and Fig.~\ref{fig:exp_setup}(a) lies in the absence of a microwave resonant cavity in Fig.~\ref{fig:exp_setup}(b), where the atomic vapor cell directly senses the microwave field radiated from a horn antenna. The difference between Fig.~\ref{fig:exp_setup}(a) and Fig.~\ref{fig:exp_setup}(c) is that, in Fig.~\ref{fig:exp_setup}(a), the microwave field is directly injected through the SMA connector from a signal generator, whereas in Fig.~\ref{fig:exp_setup}(c), two horn antennas are introduced—one for transmitting and one for receiving. The transmitting horn antenna emits a weak signal field. The receiving horn antenna, placed 1 meter away, captures this weak signal, which is then combined with a strong LO field using a power combiner before being fed into the resonant cavity. In Figs.~\ref{fig:exp_setup}(a) and \ref{fig:exp_setup}(c), the vapor cell is placed inside a metal microwave resonant cavity made of copper (see Fig.~\ref{fig:exp_setup}(e) for a photograph). The cavity is a rectangular cuboid with a length of 110\,mm, a height of 60\,mm, and a width of 55\,mm. Two small apertures, each with a diameter of 2\,mm, are opened on the movable metal plates at both ends of the resonant cavity to allow laser beams to pass through.This design, however, reduces the cavity's \(Q\)-factor. A tuning pad is installed at the center of the sidewall of the resonant cavity to adjust the resonant frequency after the cavity is mounted.  Achieving high sensitivity requires efficient coupling between the Rydberg atoms and the cavity mode, which depends on factors such as atom-cavity detuning, cavity mode volume, and the positioning of the atomic ensemble relative to the cavity field. To maximize the coupling strength, we deliberately position the atomic cell at the cavity center.We performed electromagnetic simulations using the HFSS software to optimize the atom–microwave interaction. The electric field distribution was designed to be strongly concentrated in the central region of the cavity, precisely where the atomic vapor cell is placed, as illustrated in Fig.~\ref{HFSScavity}(a). 

Furthermore, the direction of the electric field was engineered to be globally uniform across the cavity, ensuring that the polarizations of the probe laser, coupling laser, and the microwave field remain aligned for efficient optical readout, a uniformity confirmed in Fig.~\ref{HFSScavity}(b). However, the sensitivity of Rydberg atom-based microwave sensing depends not only on qualitative uniformity but also critically on quantitative alignment between the microwave electric field polarization and the atomic transition dipole moment. The Rydberg transition used in this work is \(|57D_{5/2}\rangle \rightarrow |58P_{3/2}\rangle\), whose electric dipole moment \(\mu_{34}\) has a well-defined quantization axis defined bay the polarization direction of the copropagating probe and coupling lasers. From the HFSS simulation shown in Fig.~\ref{HFSScavity}(a), the dominant polarization component of the microwave electric field in the central region (where the vapor cell is located) is along the vertical direction. In our experiment, we deliberately aligned the linear polarizations of the probe and coupling lasers to be parallel to this vertical direction. The cavity was designed to exhibit good impedance matching and high field enhancement, with a simulated $S_{11}$ of $-23.8$\,dB and an unloaded quality factor of \(Q_{0,\text{th}} = 2407.5\) at 3.778\,GHz (shown as Fig.~\ref{HFSScavity}(c)). In the actual experiment, however, the measured $S_{11}$ parameter deteriorated to $-17$\,dB, and the unloaded quality factor dropped to \(Q_{0,\text{exp}} = 616.2\). With a coupling coefficient of \(\beta_c = 0.85\), the corresponding loaded quality factors are \(Q_{L,\text{th}} \approx 1301\) and \(Q_{L,\text{exp}} \approx 333.1\), respectively. Using Eq.~(\ref{eq:sensitivity_enhance}), the theoretical enhancement factor based on the simulated \(Q_{0,\text{th}}\) is approximately \(30.7\) (29.7 dB), while the enhancement factor based on the experimentally measured \(Q_{0,\text{exp}}\) is approximately \(15.5\) (23.8 dB). 

All these discrepancies including the degraded $S_{11}$, the reduced $Q$ factor share a common physical origin. They arise from practical non-ideal factors not fully captured by the idealized simulation model, such as fabrication tolerances in the cavity dimensions, surface roughness and finite conductivity of the oxygen-free copper material, imperfections in the SMA connector soldering and assembly, as well as the presence of the two 3-mm-diameter laser access apertures on the cavity end plates. Additional insertion losses in the antenna-to-cavity coupling path, minor detuning from the exact cavity resonance, and imperfect alignment of the vapor cell relative to the field antinode further contribute to the remaining discrepancy.

\begin{figure}[t]
    \centering
    \includegraphics[width=1\linewidth]{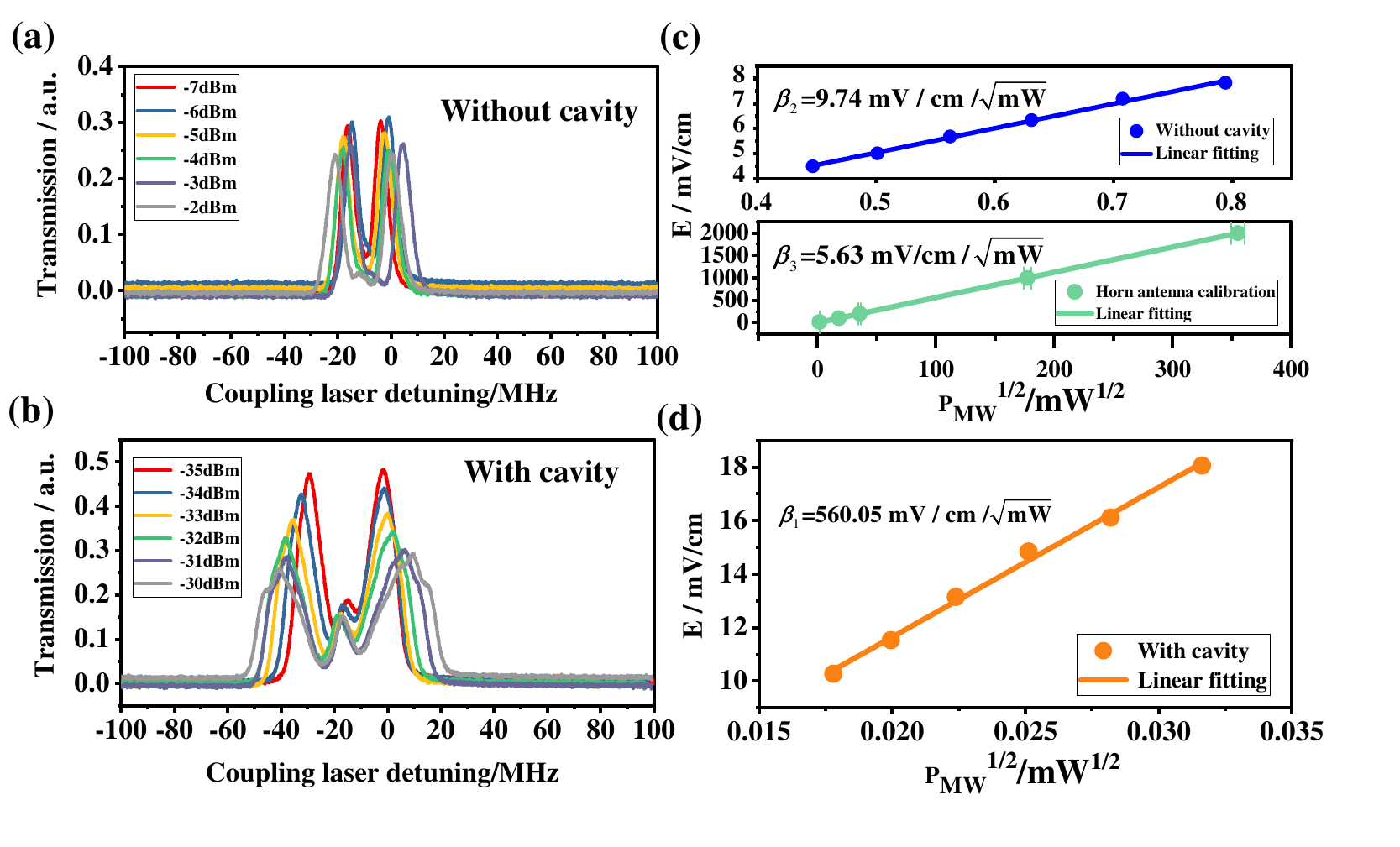}
    \caption{(a) Bare cell EIT-AT splitting (Fig.~\ref{fig:exp_setup}(b)), SIG: -7 dBm to -2 dBm. (b) Cavity direct injection EIT-AT splitting (Fig.~\ref{fig:exp_setup}(a)), SIG: -35 dBm to -30 dBm. (c) Calibration curves: upper panel shows free-space configuration (bare cell) with \(\beta_2 = 9.74\ \mathrm{mV/cm/\sqrt{mW}}\); lower panel shows antenna-coupled cavity configuration (Fig.~\ref{fig:exp_setup}(c)) with \(\beta_3 = 5.63\ \mathrm{mV/cm/\sqrt{mW}}\). (d) Direct cavity injection calibration (Fig.~\ref{fig:exp_setup}(a)): \(\beta_1 = 560.05\ \mathrm{mV/cm/\sqrt{mW}}\).}
    \label{EIT-AT}
\end{figure}

\begin{figure}[t]
    \centering
    \includegraphics[width=1\linewidth]{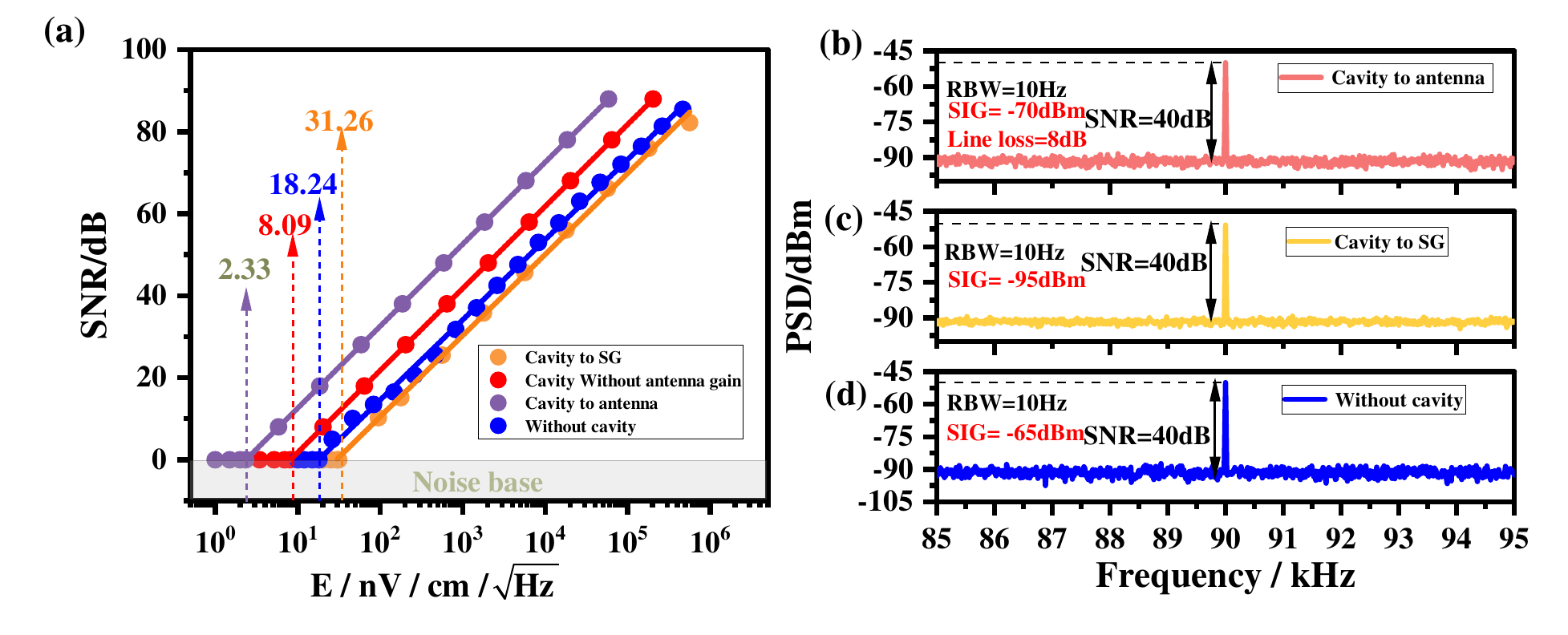}
    \caption{(a) Comparison of electric field sensitivity for three configurations: free-space (blue points with linear fit, sensitivity 18.23 nV/cm/$\sqrt{\text{Hz}}$), direct cavity injection (orange points with linear fit, sensitivity 31.84 nV/cm/$\sqrt{\text{Hz}}$), antenna-coupled cavity with antenna gain included (gray points with linear fit, sensitivity 2.33 nV/cm/$\sqrt{\text{Hz}}$), and antenna-coupled cavity without antenna gain (red points with linear fit, sensitivity 8.09 nV/cm/$\sqrt{\text{Hz}}$). (b) Spectrum of the antenna-coupled cavity configuration at SIG power of $-70$\,dBm. (c) Spectrum of the direct cavity injection configuration at SIG power of $-95$\,dBm. (d) Spectrum of the free-space configuration at SIG power of $-65$\,dBm.}
    \label{sen_com}
\end{figure}

\section{Results and Discussion}
As shown in Fig.~\ref{EIT-AT}, the AT splitting of the Rydberg EIT spectrum is presented under two different configurations. Fig.~\ref{EIT-AT}(a) corresponds to the case of a bare cesium vapor cell exposed directly to the microwave field (corresponding to Fig.~\ref{fig:exp_setup}(b)), showing the AT splitting as the SIG microwave power increases from -7 dBm to -2 dBm. Fig.~\ref{EIT-AT}(b) shows the AT splitting when the vapor cell is placed inside the microwave resonant cavity (corresponding to Fig.~\ref{fig:exp_setup}(a)), with the SIG microwave power ranging from -35 dBm to -30 dBm. Owing to the field-enhancing effect of the resonant cavity, a larger AT splitting can be achieved with a much weaker applied SIG field compared to the bare-cell configuration. In other words, the cavity allows for a well-resolved EIT-AT splitting at input microwave powers that would otherwise be insufficient to produce any clear splitting in the bare-cell case.

The relationship between the local microwave electric field strength experienced by the atoms and the applied microwave power can be described by:
\begin{equation}
E = \beta \sqrt{P_{\mathrm{MW}}},
\label{eq:calibration}
\end{equation}
where \(\beta\) is the calibration coefficient. Fig.~\ref{EIT-AT}(c) (upper panel) presents the calibration curve for the free-space (bare cell) configuration, showing this relationship with a calibration coefficient of \(\beta_2 = 9.74\ \mathrm{mV/cm/\sqrt{mW}}\). The lower panel of Fig.~\ref{EIT-AT}(c) shows the calibration curve for the antenna-coupled cavity configuration (Fig.~\ref{fig:exp_setup}(c)), illustrating the relationship between the electric field at a distance of 1 m from the transmitting horn antenna and the square root of the input power fed into the antenna, yielding \(\beta_3 = 5.63\ \mathrm{mV/cm/\sqrt{mW}}\). Fig.~\ref{EIT-AT}(d) presents the calibration curve for the direct cavity injection configuration (Fig.~\ref{fig:exp_setup}(a)), showing the relationship between the local microwave electric field strength experienced by the atoms and the square root of the applied SIG microwave power, with a calibration coefficient \(\beta_1 = 560.05\ \mathrm{mV/cm/\sqrt{mW}}\). This significant enhancement in \(\beta\) from \(\beta_2\) to \(\beta_1\) quantitatively demonstrates the cavity-induced improvement in local field strength, directly contributing to the overall sensitivity enhancement of the microwave electric field sensing.

We evaluate the electric field sensing performance under the configuration shown in Fig.~\ref{fig:exp_setup}(a), where the microwave field is directly coupled into the resonant cavity via an SMA interface. The corresponding measurement results are presented as orange points with linear fit in Fig.~\ref{sen_com}(a). Fig.~\ref{sen_com}(c) shows a representative spectrum acquired under this configuration at a SIG power of $-95$\,dBm, where the clear peak above the noise floor enables accurate sensitivity determination. In this experiment, the LO field is set to a frequency of 3.7807\,GHz, while the SIG field is detuned by 90\,kHz relative to the LO, i.e., at 3.7807\,GHz + 90\,kHz. The power of the LO field is optimized to \(-50.7\)\,dBm. The resolution bandwidth (RBW) of the spectrum analyzer is configured to 10\,Hz. The electric field sensitivity is calculated as:
\begin{equation}
E_{\mathrm{sen}} = \sqrt{10^{((P_{\mathrm{SIG}} - \mathrm{SNR} - \mathrm{RBW})/10)}} \times \beta,
\label{eq:sensitivity_direct}
\end{equation}
where \(P_{\mathrm{SIG}}\) is the signal power, SNR is the signal-to-noise ratio, and \(\beta\) is the calibration coefficient. Using this approach (see Eq.~\ref{eq:sensitivity_direct}), the sensitivity of the Rydberg-atom-based electric field sensor under this cavity-coupled configuration is determined to be \(31.84\,\text{nV/cm/}\sqrt{\text{Hz}}\).

We next examine the sensing performance under the configuration illustrated in Fig.~\ref{fig:exp_setup}(c), where the microwave fields are radiated by a transmitting horn antenna, propagate through free space over a distance of 1\,m, and are collected by a receiving horn antenna before being fed into the resonant cavity. The results with the receiving horn antenna gain included are shown as gray points with linear fit in Fig.~\ref{sen_com}(a), while the results without the receiving horn antenna gain (the antenna power gain is 10.8\,dB, corresponding to an electric field gain of 5.4\,dB) are shown as red points in Fig.~\ref{sen_com}(a). Fig.~\ref{sen_com}(b) presents a typical spectrum for this antenna-coupled cavity configuration at a SIG power of $-70$\,dBm, demonstrating a strong signal-to-noise ratio that enables high-sensitivity detection. According to the specification of the horn antenna used in the experiment, the relationship between the electric field at a distance of 1\,m from the antenna aperture and the input power fed into the antenna within the frequency band of 3.5\,GHz to 4.0\,GHz yields a calibration coefficient \(\beta_3 = 5.63\ \text{mV/cm/}\sqrt{\text{mW}}\) (as shown in lower panel of Fig.~\ref{EIT-AT}(c)). In the antenna-coupled cavity experiment, the LO and SIG frequencies remain the same as in the direct-feed case (LO at 3.7807\,GHz and SIG at 3.7807\,GHz + 90\,kHz), and the LO power is again optimized to \(-50.7\)\,dBm at the generator output. The total cable loss is measured to be \(-8\)\,dB, which is compensated when calculating the actual power delivered to the transmitting antenna. The spectrum analyzer RBW is kept at 10\,Hz. In this configuration, the electric field sensitivity is calculated as:
\begin{equation}
E_{\mathrm{sen}} = \sqrt{10^{((P_{\mathrm{SIG}} - \mathrm{Line\_loss} - \mathrm{SNR} - \mathrm{RBW})/10)}} \times \beta_3
\label{eq:sensitivity_antenna}
\end{equation}
Using Eq.~\ref{eq:sensitivity_antenna}, the resulting sensitivity is found to be \(2.33\ \text{nV/cm/}\sqrt{\text{Hz}}\). If the receiving antenna gain is removed, the sensitivity becomes \(8.09\ \text{nV/cm/}\sqrt{\text{Hz}}\).

We further evaluate the electric field sensing performance under the configuration shown in Fig.~\ref{fig:exp_setup}(b), which represents the most common experimental scheme in Rydberg-atom-based microwave sensing. The results are shown as blue points with linear fit in Fig.~\ref{sen_com}(a). Fig.~\ref{sen_com}(d) displays a representative spectrum for this free-space configuration at a SIG power of $-65$\,dBm, where the signal peak is clearly resolvable under the optimized conditions. In this setup, no microwave resonant cavity is used; instead, the microwave fields (LO and SIG) are radiated directly from a horn antenna and interact with the bare cesium atomic vapor cell. To achieve the optimal sensitivity in this configuration, we carefully optimize the laser beam sizes: the 852\,nm probe beam diameter is set to 1400\,\(\mu\)m, and the 509\,nm coupling beam diameter is set to 1500\,\(\mu\)m. Under these optimized conditions, the optimal electric field sensitivity is determined to be 18.23 nV/cm/\(\sqrt{\text{Hz}}\). This sensitivity value serves as a baseline for comparison with the antenna-coupled cavity configuration discussed above. Recall that in the cavity-enhanced scheme (Fig.~\ref{fig:exp_setup}(c)), we achieved a sensitivity of 2.33 nV/cm/\(\sqrt{\text{Hz}}\). The integration of the resonant cavity with the horn antenna feed provides an enhancement of approximately 17.9 dB in sensitivity compared to the optimized bare vapor cell configuration.The physical origin of this enhancement can be attributed to the role of the resonant cavity. In the bare vapor cell configuration, the microwave field propagates as a free-space wave, and the interaction between the Rydberg atoms and the microwave field is limited by the relatively low field strength available at the atomic position. In contrast, when the atomic vapor cell is placed inside a microwave resonant cavity, the cavity confines and concentrates the electromagnetic energy in a small mode volume. For a cavity operating near resonance, the local microwave field amplitude is significantly enhanced compared to the incident field, by a factor proportional to the square root of the cavity's quality factor \(\sqrt{Q}\). This enhanced local field increases the Rabi frequency \(\Omega_{\text{MW}}\) of the Rydberg transition, leading to a larger AT splitting for a given input microwave power. Consequently, the minimum detectable electric field is reduced. Furthermore, by positioning the vapor cell at the cavity center, we ensure that the atoms experience the maximum electric field antinode of the cavity mode, providing better mode matching between the microwave field and the atomic ensemble.

To quantitatively validate this physical picture, we compare the experimentally observed enhancement of 17.9 dB with the theoretical prediction based on the measured cavity parameters. As derived in Section~2, the sensitivity enhancement factor is given by \(\beta \sqrt{Q_{\mathrm{eff}}}\), which yields a theoretical enhancement of approximately 23.8 dB using the experimentally measured cavity parameters. This theoretical estimate exceeds the experimentally observed enhancement of 17.9 dB by approximately 5.9 dB, which can be attributed to practical non-ideal factors such as insertion losses, cavity detuning, and spatial misalignment, as discussed in Section~3. Nevertheless, the reasonable agreement between the theoretical prediction and the experimental result confirms that the observed sensitivity enhancement predominantly originates from the \(\beta \sqrt{Q_{\mathrm{eff}}}\) scaling, validating that the resonant cavity enhances the atom-microwave interaction through field confinement and resonant amplification.To further benchmark the performance of our proposed scheme, we systematically compare our experimental results with other Rydberg atom microwave receiver works reported in recent years, as summarized in Table~\ref{tab:comparison}.

\begin{table}[htbp]
\centering
\caption{Performance comparison between this work and recent representative state-of-the-art Rydberg atom microwave receivers.}
\label{tab:comparison}
\footnotesize
\begin{tabular*}{1\textwidth}{@{\extracolsep{\fill}}llll@{}}
\hline\hline
Research group (Ref.) & Frequency/GHz & Enhancement method & Sensitivity/ nV/cm/$\sqrt{\text{Hz}}$ \\
\hline
\cite{Ding2022} & 16.6  & Many-body critical enhanced  & 49  \\
\cite{Liang2026} & 2.5  & Exceptional point-Enhanced  & 22.63  \\
\cite{Wu2025} & 8.57  & laser array & 19  \\
\cite{Sandidge2024} & 10.22  & Cavity-enhanced & 15.8  \\
\cite{Yan2025} & 7.97  & Sagnac-Enhanced & 10.7  \\
\cite{Tu2024} & 36.9  & Cold atom & 10  \\
\cite{Borowka2024} & 13.9  & Six-wave mixing & 3.98  \\
\textbf{This work} & \textbf{3.78 } & \textbf{Horn-antenna-coupled cavity} & \textbf{2.33} \\
\hline\hline
\end{tabular*}
\end{table}

\section{Conclusion}
In summary, we have experimentally demonstrated a resonant microwave cavity approach to significantly enhance the sensitivity of a cesium-based Rydberg atom receiver operating in the S-band. By systematically comparing three configurations—bare vapor cell, direct cavity injection, and antenna-coupled cavity—we have quantified the sensitivity improvement enabled by the cavity. The bare vapor cell configuration, with optimized laser beam sizes and cell temperature, yields a baseline sensitivity of 18.23 nV/cm/$\sqrt{\text{Hz}}$. Direct cavity injection via an SMA interface gives a sensitivity of 31.84 nV/cm/$\sqrt{\text{Hz}}$, which is limited by the available microwave power coupling. Most importantly, the antenna-coupled cavity configuration, which emulates realistic free-space wireless reception over a distance of 1 m, achieves a sensitivity of 2.33 nV/cm/$\sqrt{\text{Hz}}$ at the receiving antenna, corresponding to an enhancement of approximately 17.9 dB compared to the optimized bare vapor cell. This enhancement arises from the combined effects of cavity field confinement, resonant field amplification, and optimized spatial overlap between the atomic ensemble and the cavity mode. Our work provides a practical and effective route for deploying high-sensitivity Rydberg atom-based microwave receivers in real-world communication and sensing applications, where weak free-space microwave signals must be reliably detected.
\section*{Declaration of competing interest}
The authors declare that they have no known competing financial interests or personal relationships that could have appeared to influence the work reported in this paper.
\section*{Acknowledgements}

This work was supported by the National Key Research and Development Program of China (Grant No. 2024YFB3909500), the National Natural Science Foundation of China (Grant No. U2341211), the National Natural Science Foundation of China (Grant No. 62501568).
\section*{Data availability}
Data will be made available on request.
\bibliographystyle{unsrt}
\bibliography{sample}

@ARTICLE{Lei2026,
  author={Lei, Mingwei and Shi, Meng},
  journal={IEEE Transactions on Quantum Engineering}, 
  title={Satellite Microwave Detection via Cavity-Coupled Rydberg Atomic Receiver}, 
  year={2026},
  volume={},
  number={},
  pages={1-6},
  doi={10.1109/TQE.2026.3690617}}

@article{ Arumugam2024,
Author = {Arumugam, Darmindra and Park, Jun-Hee and Feyissa, Brook and Bush, Jack
   and Mysore Nagaraja, Srinivas Prasad},
Title = {Remote sensing of soil moisture using Rydberg atoms and satellite
   signals of opportunity},
Journal = {SCIENTIFIC REPORTS},
Year = {2024},
Volume = {14},
Number = {1},
Month = {AUG 4},
DOI = {10.1038/s41598-024-68914-6},
Article-Number = {18025},
ISSN = {2045-2322},
Unique-ID = {WOS:001283964800019},
}

@article{ Huang2025,
Author = {Huang, Jingyi and Sehgal, Vinit and Alvarez, Laura V. and Brocca, Luca
   and Cai, Shuohao and Cheng, Rui and Cheng, Xinghua and Du, Jinyang and
   El Masri, Bassil and Endsley, K. Arthur and Fang, Yilin and Hu, Jie and
   Jampani, Mahesh and Kibria, Md Golam and Koren, Gerbrand and Li,
   Lingcheng and Liu, Laibao and Mao, Jiafu and Moreno, Hernan A. and
   Rigden, Angela and Shi, Mingjie and Shi, Xiaoying and Wang, Yaoping and
   Zhang, Xi and Fisher, Joshua B.},
Title = {Remotely Sensed High-Resolution Soil Moisture and Evapotranspiration:
   Bridging the Gap Between Science and Society},
Journal = {WATER RESOURCES RESEARCH},
Year = {2025},
Volume = {61},
Number = {5},
Month = {MAY 13},
DOI = {10.1029/2024WR037929},
Article-Number = {e2024WR037929},
ISSN = {0043-1397},
EISSN = {1944-7973},
ResearcherID-Numbers = {Huang, Jingyi/B-5541-2015
   Fisher, Joshua/AFM-8914-2022
   Álvarez-Álvarez, Laura/AHC-8249-2022
   Cheng, Rui/HGD-0865-2022
   Li, Lingcheng/HTT-4820-2023
   Brocca, Luca/F-2854-2010
   CHENG, Xinghua/KDO-2940-2024
   Mao, Jiafu/B-9689-2012
   Moreno, Hernan/H-2337-2014
   Hu, Jie/JQI-6149-2023
   Jampani, Mahesh/S-8682-2018
   Koren, Gerbrand/IZE-2064-2023
   Shi, Mingjie/LFS-4086-2024
   Liu, Laibao/MEP-5116-2025
   Fang, Yilin/J-5137-2015
   du, jinyang/LZE-5392-2025
   Zhang, Xi/ABB-8182-2020},
ORCID-Numbers = {Huang, Jingyi/0000-0002-1209-9699
   Wang, Yaoping/0000-0002-5162-1910
   Kibria, Golam/0009-0004-3448-3589
   Cheng, Rui/0000-0002-3003-8339
   Brocca, Luca/0000-0002-9080-260X
   CHENG, Xinghua/0000-0002-1931-6174
   Alvarez, Laura V/0000-0002-5047-5384
   Cai, Shuohao/0009-0007-7114-1395
   Jampani, Mahesh/0000-0002-8925-719X
   El Masri, Bassil/0000-0002-1017-5467
   Endsley, K. Arthur/0000-0001-9722-8092
   Fang, Yilin/0000-0003-1969-9889
   Zhang, Xi/0000-0002-6519-4569},
Unique-ID = {WOS:001487825500001},
}

@article{ Sun2025,
Author = {Sun, Wei and Chen, Zhiyang and Li, Linze and Shen, Chen and Yu, Kunpeng
   and Li, Shichang and Long, Jinbao and Zheng, Huamin and Wang, Luyu and
   Long, Tianyu and Chen, Qiushi and Zhang, Zhouze and Shi, Baoqi and Gao,
   Lan and Luo, Yi-Han and Chen, Baile and Liu, Junqiu},
Title = {A chip-integrated comb-based microwave oscillator},
Journal = {LIGHT-SCIENCE \& APPLICATIONS},
Year = {2025},
Volume = {14},
Number = {1},
Month = {APR 30},
DOI = {10.1038/s41377-025-01795-0},
Article-Number = {179},
ISSN = {2095-5545},
EISSN = {2047-7538},
ResearcherID-Numbers = {Wang, Luyu/OXC-5621-2025
   Li, Linze/J-5187-2017
   Baoqi, Shi/HPH-7041-2023
   Liu, Junqiu/AAT-2765-2021},
ORCID-Numbers = {Wang, Luyu/0000-0001-9846-6607
   Chen, Qiushi/0009-0009-0898-9734
   Long, JinBao/0009-0000-9341-1113
   },
Unique-ID = {WOS:001478992400001},
}

@article{ Tan2025,
Author = {Tan, Xin and Gu, Weihua and Tao, Zhe and Chen, Xiangling and Li, Siyuan
   and Xia, Ailin and Ji, Guangbin},
Title = {Carbon-based materials for radar-infrared compatible stealth technology},
Journal = {CHEMICAL ENGINEERING JOURNAL},
Year = {2025},
Volume = {507},
Month = {MAR 1},
DOI = {10.1016/j.cej.2025.160168},
EarlyAccessDate = {FEB 2025},
Article-Number = {160168},
ISSN = {1385-8947},
EISSN = {1873-3212},
ResearcherID-Numbers = {Gu, Weihua/ABG-5395-2021
   , Ji/AAZ-4756-2020},
ORCID-Numbers = {, Ji/0000-0001-5782-1588},
Unique-ID = {WOS:001427238800001},
}

@article{ Fan2025,
Author = {Fan, Dong and Zhao, Tianjie and Jiang, Xiaoguang and Garcia-Garcia,
   Almudena and Schmidt, Toni and Samaniego, Luis and Attinger, Sabine and
   Wu, Hua and Jiang, Yazhen and Shi, Jiancheng and Fan, Lei and Tang,
   Bo-Hui and Wagner, Wolfgang and Dorigo, Wouter and Gruber, Alexander and
   Mattia, Francesco and Balenzano, Anna and Brocca, Luca and Jagdhuber,
   Thomas and Wigneron, Jean-Pierre and Montzka, Carsten and Peng, Jian},
Title = {A Sentinel-1 SAR-based global 1-km resolution soil moisture data
   product: Algorithm and preliminary assessment},
Journal = {REMOTE SENSING OF ENVIRONMENT},
Year = {2025},
Volume = {318},
Month = {MAR 1},
DOI = {10.1016/j.rse.2024.114579},
EarlyAccessDate = {JAN 2025},
Article-Number = {114579},
ISSN = {0034-4257},
EISSN = {1879-0704},
ResearcherID-Numbers = {Fan, Lei/LQK-1265-2024
   Samaniego Eguiguren, Luis Eduardo/G-8651-2011
   Wigneron, Jean-Pierre/ABD-9939-2021
   Samaniego, Luis/G-8651-2011
   Gruber, Alexander/ABA-3286-2020
   Brocca, Luca/F-2854-2010
   Wagner, Wolfgang/AAC-5507-2019
   Attinger, Sabine/LWH-7754-2024
   Dorigo, Wouter/C-7794-2014
   García-García, Almudena/AGW-0891-2022
   Peng, Jian/IQT-8432-2023
   Montzka, Carsten/D-1617-2009},
ORCID-Numbers = {Schmidt, Toni/0000-0002-2134-7527
   Balenzano, Anna/0000-0002-9355-9035
   Samaniego Eguiguren, Luis Eduardo/0000-0002-8449-4428
   fan, dong/0000-0002-4604-176X
   Wagner, Wolfgang/0000-0001-7704-6857
   Peng, Jian/0000-0002-4071-0512
   },
Unique-ID = {WOS:001411113400001},
}

@article{ Saad2020,
Author = {Saad, Walid and Bennis, Mehdi and Chen, Mingzhe},
Title = {A Vision of 6G Wireless Systems: Applications, Trends, Technologies, and
   Open Research Problems},
Journal = {IEEE NETWORK},
Year = {2020},
Volume = {34},
Number = {3},
Pages = {134-142},
Month = {MAY-JUN},
DOI = {10.1109/MNET.001.1900287},
ISSN = {0890-8044},
EISSN = {1558-156X},
ResearcherID-Numbers = {Saad, Walid/C-7978-2018
   Chen, Mingzhe/U-3377-2019
   Bennis, Mehdi/ABE-5838-2020},
ORCID-Numbers = {Chen, Mingzhe/0000-0003-2570-703X
   Bennis, Mehdi/0000-0003-0261-0171},
Unique-ID = {WOS:000541150100019},
}

@article{ Rappaport2013,
Author = {Rappaport, Theodore S. and Sun, Shu and Mayzus, Rimma and Zhao, Hang and
   Azar, Yaniv and Wang, Kevin and Wong, George N. and Schulz, Jocelyn K.
   and Samimi, Mathew and Gutierrez, Felix},
Title = {Millimeter Wave Mobile Communications for 5G Cellular: It Will Work!},
Journal = {IEEE ACCESS},
Year = {2013},
Volume = {1},
Pages = {335-349},
DOI = {10.1109/ACCESS.2013.2260813},
ISSN = {2169-3536},
ResearcherID-Numbers = {Wong, George/AAL-1016-2021
   Sun, Shu/MFJ-3517-2025
   Rodero, Félix/Q-7880-2018},
ORCID-Numbers = {Wong, George/0000-0001-6952-2147
   Rappaport, Theodore/0000-0001-7449-9957
   },
Unique-ID = {WOS:000209652700025},
}

@article{ Liang2026,
Author = {Liang, Chao and Yang, Ce and Huang, Wei and You, Li},
Title = {Exceptional Point-Enhanced Rydberg Atomic Electrometers},
Journal = {PHYSICAL REVIEW LETTERS},
Year = {2026},
Volume = {136},
Number = {5},
Month = {FEB 5},
DOI = {10.1103/jptr-pm37},
Article-Number = {053203},
ISSN = {0031-9007},
EISSN = {1079-7114},
Unique-ID = {WOS:001686761400004},
}

@article{ Ding2022,
Author = {Ding, Dong-Sheng and Liu, Zong-Kai and Shi, Bao-Sen and Guo, Guang-Can
   and Molmer, Klaus and Adams, Charles S.},
Title = {Enhanced metrology at the critical point of a many-body Rydberg atomic
   system},
Journal = {NATURE PHYSICS},
Year = {2022},
Volume = {18},
Number = {12},
Month = {DEC},
DOI = {10.1038/s41567-022-01777-8},
EarlyAccessDate = {OCT 2022},
ISSN = {1745-2473},
EISSN = {1745-2481},
ResearcherID-Numbers = {Moelmer, Klaus/KRO-7723-2024
   Ada, Charles/C-8808-2015
   },
ORCID-Numbers = {Ding, Dong-Sheng/0000-0002-5051-4777
   Moelmer, Klaus/0000-0002-2372-869X
   Ada, Charles/0000-0001-5602-2741
   Liu, ZongKai/0000-0001-8851-0715},
Unique-ID = {WOS:000868967000001},
}

@article{ Wu2025,
Author = {Wu, Bo and Mao, Ruiqi and Sang, Di and Sun, Zhanshan and Liu, Yi and
   Lin, Yi and An, Qiang and Fu, Yunqi},
Title = {Enhancing Sensitivity of Atomic Microwave Receivers Based on Optimal
   Laser Arrays},
Journal = {IEEE TRANSACTIONS ON ANTENNAS AND PROPAGATION},
Year = {2025},
Volume = {73},
Number = {2},
Pages = {793-806},
Month = {FEB},
DOI = {10.1109/TAP.2024.3486553},
ISSN = {0018-926X},
EISSN = {1558-2221},
ResearcherID-Numbers = {Lin, Yi/MXJ-7360-2025
   Liu, Yi/N-1184-2015
   },
ORCID-Numbers = {Lin, Yi/0000-0002-9235-3668
   Liu, Yi/0000-0002-2558-444X
   Sang, Di/0000-0003-1247-605X
   , Zhanshan Sun/0000-0003-2918-7502
   Mao, Ruiqi/0000-0002-9803-9784},
Unique-ID = {WOS:001414904400031},
}

@article{ Yan2025,
Author = {Yan, Hongmei and Gao, Taisen and Jing, Mingyong and Yang, Weguang and
   Zhang, Hao and Liu, Zongkai and Xie, Junyao and Xiao, Liantuan and Jia,
   Suotang and Zhang, Linjie},
Title = {Sagnac-Enhanced Rydberg Superheterodyne Receiver with Dual-Beam
   Interference},
Journal = {CHINESE PHYSICS LETTERS},
Year = {2025},
Volume = {42},
Number = {12},
Month = {DEC 1},
DOI = {10.1088/0256-307X/42/12/120601},
Article-Number = {120601},
ISSN = {0256-307X},
EISSN = {1741-3540},
Unique-ID = {WOS:001631321900001},
}

@article{ Tu2024,
Author = {Tu, Hai-Tao and Liao, Kai-Yu and Wang, Hong-Lei and Zhu, Yi-Fei and Qiu,
   Si-Yuan and Jiang, Hao and Huang, Wei and Bian, Wu and Yan, Hui and Zhu,
   Shi-Liang},
Title = {Approaching the standard quantum limit of a Rydberg-atom microwave
   electrometer},
Journal = {SCIENCE ADVANCES},
Year = {2024},
Volume = {10},
Number = {51},
Month = {DEC 20},
DOI = {10.1126/sciadv.ads0683},
Article-Number = {eads0683},
ISSN = {2375-2548},
ResearcherID-Numbers = {qiu, siyuan/OZF-4215-2025
   Zhu, Shi-Liang/F-2334-2011
   Yan, Hui/I-8294-2014},
ORCID-Numbers = {Bian, Wu/0009-0004-2428-8069
   Wang, Hong-Lei/0009-0002-5499-0022
   , Wei/0000-0001-8021-1665
   Zhu, Shi-Liang/0000-0002-8913-9847
   Tu, Hai-Tao/0000-0001-9920-1870
   Liao, Kai-Yu/0000-0002-4577-7150
   Yan, Hui/0000-0002-7858-2368},
Unique-ID = {WOS:001381247200021},
}

@article{ Robinson2021,
Author = {Robinson, Amy K. and Prajapati, Nikunjkumar and Senic, Damir and Simons,
   Matthew T. and Holloway, Christopher L.},
Title = {Determining the angle-of-arrival of a radio-frequency source with a
   Rydberg atom-based sensor},
Journal = {APPLIED PHYSICS LETTERS},
Year = {2021},
Volume = {118},
Number = {11},
Month = {MAR 15},
DOI = {10.1063/5.0045601},
Article-Number = {114001},
ISSN = {0003-6951},
EISSN = {1077-3118},
ORCID-Numbers = {Prajapati, Nikunj/0000-0002-7779-9741
   Robinson, Amy/0000-0002-6505-1920},
Unique-ID = {WOS:000629818300001},
}

@article{ Holloway2019-2,
Author = {Holloway, Christopher L. and Simons, Matthew T. and Haddab, Abdulaziz H.
   and Williams, Carl J. and Holloway, Maxwell W.},
Title = {A ``real-time{''} guitar recording using Rydberg atoms and
   electromagnetically induced transparency: Quantum physics meets music},
Journal = {AIP ADVANCES},
Year = {2019},
Volume = {9},
Number = {6},
Month = {JUN},
DOI = {10.1063/1.5099036},
Article-Number = {065110},
ISSN = {2158-3226},
ResearcherID-Numbers = {Haddab, Abdulaziz/AAX-9569-2020
   Willia, Carl/B-5877-2009
   Holloway, Maxwell/AAB-1293-2021},
ORCID-Numbers = {SIMONS, MATTHEW/0000-0001-9418-7520
   Haddab, Abdulaziz/0000-0003-2803-6091
   Willia, Carl/0000-0003-4274-3227
   },
Unique-ID = {WOS:000474430700042},
}

@article{ Holloway2019,
Author = {Holloway, Christopher L. and Simons, Matthew T. and Gordon, Joshua A.
   and Novotny, David},
Title = {Detecting and Receiving Phase-Modulated Signals With a Rydberg
   Atom-Based Receiver},
Journal = {IEEE ANTENNAS AND WIRELESS PROPAGATION LETTERS},
Year = {2019},
Volume = {18},
Number = {9},
Pages = {1853-1857},
Month = {SEP},
DOI = {10.1109/LAWP.2019.2931450},
ISSN = {1536-1225},
EISSN = {1548-5757},
ORCID-Numbers = {SIMONS, MATTHEW/0000-0001-9418-7520},
Unique-ID = {WOS:000489761700031},
}

@article{ Jiao2019,
Author = {Jiao, Yuechun and Han, Xiaoxuan and Fan, Jiabei and Raithel, Georg and
   Zhao, Jianming and Jia, Suotang},
Title = {Atom-based receiver for amplitude-modulated baseband signals in
   high-frequency radio communication},
Journal = {APPLIED PHYSICS EXPRESS},
Year = {2019},
Volume = {12},
Number = {12},
Month = {DEC 1},
DOI = {10.7567/1882-0786/ab5463},
Article-Number = {126002},
ISSN = {1882-0778},
EISSN = {1882-0786},
ORCID-Numbers = {Zhao, Jianming/0000-0001-8420-9319
   Raithel, Georg/0000-0002-2005-8440},
Unique-ID = {WOS:000499354200001},
}

@article{ Meyer2018,
Author = {Meyer, David H. and Cox, Kevin C. and Fatemi, Fredrik K. and Kunz, Paul
   D.},
Title = {Digital communication with Rydberg atoms and amplitude-modulated
   microwave fields},
Journal = {APPLIED PHYSICS LETTERS},
Year = {2018},
Volume = {112},
Number = {21},
Month = {MAY 21},
DOI = {10.1063/1.5028357},
Article-Number = {211108},
ISSN = {0003-6951},
EISSN = {1077-3118},
ResearcherID-Numbers = {Cox Jr, Kevin/LEN-3441-2024
   Meyer, David H/AAE-2908-2019},
ORCID-Numbers = {Meyer, David H/0000-0003-2452-2017},
Unique-ID = {WOS:000433140900008},
}

@article{ Cox2021,
Author = {Meyer, David H. and Kunz, Paul D. and Cox, Kevin C.},
Title = {Waveguide-Coupled Rydberg Spectrum Analyzer from 0 to 20 GHz},
Journal = {PHYSICAL REVIEW APPLIED},
Year = {2021},
Volume = {15},
Number = {1},
Month = {JAN 27},
DOI = {10.1103/PhysRevApplied.15.014053},
Article-Number = {014047},
ISSN = {2331-7019},
ResearcherID-Numbers = {Cox Jr, Kevin/LEN-3441-2024
   Meyer, David H/AAE-2908-2019},
ORCID-Numbers = {Meyer, David H/0000-0003-2452-2017},
Unique-ID = {WOS:000612214400006},
}

@article{ Yang2023,
Author = {Yang, Kai and Mao, Ruiqi and He, Li and Yao, Jiawei and Li, Jianbing and
   Sun, Zhanshan and Fu, Yunqi},
Title = {Local oscillator port embedded field enhancement resonator for Rydberg
   atomic heterodyne technique},
Journal = {EPJ QUANTUM TECHNOLOGY},
Year = {2023},
Volume = {10},
Number = {1},
Month = {DEC},
DOI = {10.1140/epjqt/s40507-023-00179-w},
Article-Number = {23},
ISSN = {2662-4400},
EISSN = {2196-0763},
ResearcherID-Numbers = {Yang, Kai/IXD-4311-2023
   Li, Jianbing/AAS-2476-2021
   He, Li/C-1320-2014},
Unique-ID = {WOS:001020639200002},
}

@article{ Holloway2022,
Author = {Holloway, Christopher L. and Prajapati, Nikunjkumar and Artusio-Glimpse,
   Alexandra B. and Berweger, Samuel and Simons, Matthew T. and Kasahara,
   Yoshiaki and Alu, Andrea and Ziolkowski, Richard W.},
Title = {Rydberg atom-based field sensing enhancement using a split-ring
   resonator},
Journal = {APPLIED PHYSICS LETTERS},
Year = {2022},
Volume = {120},
Number = {20},
Month = {MAY 16},
DOI = {10.1063/5.0088532},
Article-Number = {204001},
ISSN = {0003-6951},
EISSN = {1077-3118},
ResearcherID-Numbers = {Artusio-Glimpse, Alexandra/AAE-2433-2019
   Kasahara, Yoshiaki/JKH-4814-2023
   Alu, Andrea/A-1328-2007},
ORCID-Numbers = {BERWEGER, SAMUEL/0000-0002-4073-5322
   Prajapati, Nikunj/0000-0002-7779-9741
   Ziolkowski, Richard/0000-0003-4256-6902
   Alu, Andrea/0000-0002-4297-5274},
Unique-ID = {WOS:000799096000002},
}

@article{ Walker2022,
Author = {Walker, D. M. and Brown, L. L. and Hogan, S. D.},
Title = {Electrometry of a single resonator mode at a
   Rydberg-atom-superconducting-circuit interface},
Journal = {PHYSICAL REVIEW A},
Year = {2022},
Volume = {105},
Number = {2},
Month = {FEB 28},
DOI = {10.1103/PhysRevA.105.022626},
Article-Number = {022626},
ISSN = {2469-9926},
EISSN = {2469-9934},
ResearcherID-Numbers = {Hogan, Stephen/LDG-4137-2024},
ORCID-Numbers = {Brown, Luke Lister/0009-0009-3507-7909
   Hogan, Stephen/0000-0002-7720-3979},
Unique-ID = {WOS:000766647300011},
}

@article{ Hogan2012,
Author = {Hogan, S. D. and Agner, J. A. and Merkt, F. and Thiele, T. and Filipp,
   S. and Wallraff, A.},
Title = {Driving Rydberg-Rydberg Transitions from a Coplanar Microwave Waveguide},
Journal = {PHYSICAL REVIEW LETTERS},
Year = {2012},
Volume = {108},
Number = {6},
Month = {FEB 9},
DOI = {10.1103/PhysRevLett.108.063004},
Article-Number = {063004},
ISSN = {0031-9007},
ResearcherID-Numbers = {Filipp, Stefan/F-4775-2013
   Hogan, Stephen/LDG-4137-2024
   Wallraff, Aneas/C-2130-2009
   Thiele, Tobias/JCE-2737-2023},
ORCID-Numbers = {Merkt, Frederic/0000-0002-4897-2234
   Hogan, Stephen/0000-0002-7720-3979
   Wallraff, Aneas/0000-0002-3476-4485
   },
Unique-ID = {WOS:000300101500007},
}

@article{ Morgan2020,
Author = {Morgan, A. A. and Hogan, S. D.},
Title = {Coupling Rydberg Atoms to Microwave Fields in a Superconducting Coplanar
   Waveguide Resonator},
Journal = {PHYSICAL REVIEW LETTERS},
Year = {2020},
Volume = {124},
Number = {19},
Month = {MAY 12},
DOI = {10.1103/PhysRevLett.124.193604},
Article-Number = {193604},
ISSN = {0031-9007},
EISSN = {1079-7114},
ResearcherID-Numbers = {Hogan, Stephen/LDG-4137-2024
   },
ORCID-Numbers = {Hogan, Stephen/0000-0002-7720-3979
   Morgan, Alexane/0000-0003-3192-9587},
Unique-ID = {WOS:000531741700014},
}

@article{Schlossberger2024,
Author = {Schlossberger, Noah and Prajapati, Nikunjkumar and Berweger, Samuel and
   Rotunno, Andrew P. and Artusio-Glimpse, Alexandra B. and Simons, Matthew
   T. and Sheikh, Abrar A. and Norrgard, Eric B. and Eckel, Stephen P. and
   Holloway, Christopher L.},
Title = {Rydberg states of alkali atoms in atomic vapour as SI-traceable field
   probes and communications receivers},
Journal = {NATURE REVIEWS PHYSICS},
Year = {2024},
Volume = {6},
Number = {10},
Pages = {606-620},
Month = {OCT},
DOI = {10.1038/s42254-024-00756-7},
EarlyAccessDate = {SEP 2024},
EISSN = {2522-5820},
ResearcherID-Numbers = {Artusio-Glimpse, Alexandra/AAE-2433-2019
   Eckel, Stephen/K-5215-2014},
ORCID-Numbers = {Norrgard, Eric/0000-0002-8715-4648
   Schlossberger, Noah/0000-0001-9573-8152
   },
Unique-ID = {WOS:001313607700001},
}

@article{Figger1980,
  title={A photon detector for submillimetre wavelengths using Rydberg atoms},
  author={ Figger, H.  and  Leuchs, G.  and  Straubinger, R.  and  Walther, H. },
  journal={Optics Communications},
  volume={33},
  number={1},
  pages={37-41},
  year={1980},
}

@article{ Prajapati2021,
Author = {Prajapati, Nikunjkumar and Robinson, Amy K. and Berweger, Samuel and
   Simons, Matthew T. and Artusio-Glimpse, Alexandra B. and Holloway,
   Christopher L.},
Title = {Enhancement of electromagnetically induced transparency based
   Rydberg-atom electrometry through population repumping},
Journal = {APPLIED PHYSICS LETTERS},
Year = {2021},
Volume = {119},
Number = {21},
Month = {NOV 22},
DOI = {10.1063/5.0069195},
Article-Number = {214001},
ISSN = {0003-6951},
EISSN = {1077-3118},
ResearcherID-Numbers = {Artusio-Glimpse, Alexandra/AAE-2433-2019
   },
ORCID-Numbers = {Robinson, Amy/0000-0002-6505-1920
   Prajapati, Nikunj/0000-0002-7779-9741
   BERWEGER, SAMUEL/0000-0002-4073-5322},
Unique-ID = {WOS:000749549600009},
}

@article{ Cai2022,
Author = {Cai, Minghao and Xu, Zishan and You, Shuhang and Liu, Hongping},
Title = {Sensitivity Improvement and Determination of Rydberg Atom-Based
   Microwave Sensor},
Journal = {PHOTONICS},
Year = {2022},
Volume = {9},
Number = {4},
Month = {APR},
DOI = {10.3390/photonics9040250},
Article-Number = {250},
EISSN = {2304-6732},
ORCID-Numbers = {Cai, M H/0000-0002-7224-6152},
Unique-ID = {WOS:000785664700001},
}

@article{ Borowka2024,
Author = {Borowka, Sebastian and Pylypenko, Uliana and Mazelanik, Mateusz and
   Parniak, Michal},
Title = {Continuous wideband microwave-to-optical converter based on
   room-temperature Rydberg atoms},
Journal = {NATURE PHOTONICS},
Year = {2024},
Volume = {18},
Number = {1},
Month = {JAN},
DOI = {10.1038/s41566-023-01295-w},
EarlyAccessDate = {OCT 2023},
ISSN = {1749-4885},
EISSN = {1749-4893},
ResearcherID-Numbers = {Mazelanik, Mateusz/JMC-5930-2023
   Parniak, Michał/C-5765-2015},
ORCID-Numbers = {Pylypenko, Uliana/0009-0002-7560-1903
   Borówka, Sebastian/0000-0003-4085-6076
   Parniak, Michał/0000-0002-6849-4671},
Unique-ID = {WOS:001080198800001},
}

@article{ Liu2025,
Author = {Liu, Bang and Zhang, Li-Hua and Wang, Qi-Feng and Ma, Yu and Han,
   Tian-Yu and Liu, Zong-Kai and Zhang, Zheng-Yuan and Shao, Shi-Yao and
   Zhang, Jun and Li, Qing and Chen, Han-Chao and Han, Yu-Long and Ding,
   Dong-Sheng and Shi, Bao-Sen},
Title = {Cavity-Enhanced Rydberg Atom Microwave Receiver},
Journal = {CHINESE PHYSICS LETTERS},
Year = {2025},
Volume = {42},
Number = {5},
Month = {MAY 1},
DOI = {10.1088/0256-307X/42/5/053201},
Article-Number = {053201},
ISSN = {0256-307X},
EISSN = {1741-3540},
ResearcherID-Numbers = {Zhang, Jun/G-7058-2011},
Unique-ID = {WOS:001493203900001},
}

@article{ Sandidge2024,
Author = {Sandidge, Georgia and Santamaria-Botello, Gabriel and Bottomley, Eric
   and Fan, Haoquan and Popovic, Zoya},
Title = {Resonant Structures for Sensitivity Enhancement of Rydberg-Atom
   Microwave Receivers},
Journal = {IEEE TRANSACTIONS ON MICROWAVE THEORY AND TECHNIQUES},
Year = {2024},
Volume = {72},
Number = {4},
Pages = {2057-2066},
Month = {APR},
DOI = {10.1109/TMTT.2024.3355763},
EarlyAccessDate = {FEB 2024},
ISSN = {0018-9480},
EISSN = {1557-9670},
ORCID-Numbers = {Sandidge, Georgia/0009-0001-8142-1864
   Popovic, Zorana/0000-0001-7651-2254
   SANTAMARIA BOTELLO, GABRIEL/0000-0003-4736-0030},
Unique-ID = {WOS:001168603600001},
}
\end{document}